\documentclass[aps,pre,reprint,superscriptaddress,longbibliography]{revtex4-1}
\usepackage{amsmath,amssymb,graphicx,subfigure,color,times,tabularx}

\begin{document}
\title{On the relation between the velocity- and position-Verlet integrators}
\author{Liyan Ni}
\affiliation{Institute of Frontier Chemistry, School of Chemistry and Chemical Engineering, Shandong University, Qingdao,   266237,  P. R. China}
\affiliation{Qingdao Institute for Theoretical and Computational Sciences (QiTCS), Shandong University, Qingdao, 266237, P. R. China}
\author{Zhonghan Hu} \email{zhonghanhu@sdu.edu.cn}
\affiliation{Institute of Frontier Chemistry, School of Chemistry and Chemical Engineering, Shandong University, Qingdao,   266237,  P. R. China}
\affiliation{Qingdao Institute for Theoretical and Computational Sciences (QiTCS), Shandong University, Qingdao, 266237, P. R. China}
\begin{abstract}
The difference and similarity between the velocity- and position-Verlet integrators are discussed from the viewpoint of their Hamiltonian representations for both linear and nonlinear systems. 
For a harmonic oscillator, the exact Hamiltonians reveal that positional trajectories generated by the two integrators follow an identical second-order differential equation and thus can be matched by adjusting initial conditions.
In contrast, the series expansion of the Hamiltonians for the nonlinear discrete dynamics clearly indicate that the two integrators differ fundamentally.
These analytical results are confirmed by simple numerical simulations of harmonic and anharmonic oscillators. \end{abstract}
\maketitle

About three decades ago, Tuckerman {\it et al.} suggested applying the Trotter factorization of the Liouville propagator to systematically generate time-reversible symplectic molecular dynamics integrators\cite{Tuckerman1992}.
Their formulation has been widely adapted to producing novel algorithms for simulations with or without an external coupling to thermo- and baro-stats (e.g. Refs.\cite{Frenkel_Smit2023} and references therein).
In the original work\cite{Tuckerman1992}, they showed how to use the Trotter factorization to easily derive the well known velocity-Verlet integrator\cite{Swope1982}, 
which updates the position and velocity at the time $t=n\tau$, $(x(n\tau), v(n\tau))=(x_n, v_n)$ with the time-interval $\tau$ fixed and $n\geqslant 0$ an integer, according to (Ref.\cite{Tuckerman1992}, Eq.(2.18))
\begin{equation}\left\{ \begin{aligned}\displaystyle x_{n+1} & = x_n + \tau\, v_n + \dfrac{\tau^2}{2} f(x_n) \\
      v_{n+1} & = v_n +\dfrac{\tau}{2} \left[ f(x_n) + f(x_{n+1}) \right] \end{aligned} \right.\label{eq:v}, \end{equation}
where $f(x)$ is the position-dependent force reduced by mass. 
Interestingly, the Trotter factorization was also used to derive an entirely new integrator, which they named the position-Verlet integrator, updating the position and velocity, $(y(n\tau),w(n\tau))=(y_n,w_n)$, according to (Ref.\cite{Tuckerman1992}, Eq.(2.22))
\begin{equation}\left\{ \begin{aligned} y_{n+1} & = y_n + \tau\, w_n + \dfrac{\tau^2}{2} f(y_n+w_n\tau/2) \\
      w_{n+1} &= w_n +  \tau\, f(y_n +  w_n \tau/2 ) \end{aligned} \right. \label{eq:p}. \end{equation}
Eq.~\eqref{eq:p} differs from Eq.~\eqref{eq:v} in that the force is always evaluated at the position of a half time-step: $y_n+w_n\tau/2$.

By setting identical initial conditions: $x_0 = y_0$ and $v_0 = w_0$ for either the Lennard-Jones fluid\cite{Tuckerman1992} or the harmonic oscillator giving the linear force: $f(x)=-x$\cite{Tuckerman1993}, 
it was demonstrated numerically that the two integrators, Eqs.~\eqref{eq:v} and~\eqref{eq:p} outputted distinct trajectories, as expected.
However, Toxvaerd\cite{Toxvaerd1993,note2021} argued that both integrators can be reduced to the Verlet algorithm (Eq.(7) of ref.\cite{Toxvaerd1993,Farago_note}) 
\begin{equation}  r_{n+2} = 2r_{n+1} - r_n + \tau^2 f(r_{n+1}) \label{eq:ov}, \end{equation}
which, by discretizing time, approximates the Newton's equation of motion governing the continuous evolution of the position $r$ only.
In the viewpoint of Ref.\cite{Toxvaerd1993}, all formulations, i.e., Eqs.~\eqref{eq:v} to~\eqref{eq:ov} in our notation, generate discrete trajectories that identically follow the exact time evolution of a slight perturbed Hamiltonian\cite{Yoshida1992}.
Since no explicit expression for any nontrivial perturbed Hamiltonian was available at that time, it was unknown to what extent this viewpoint remains valid although it appears to be certainly informative.
Fortunately, the complete and explicit perturbed Hamiltonian for any discrete dynamics of the harmonic oscillator has now been solved exactly\cite{Ni_Hu2024} and thus allows a direct evaluation of the relation between Eqs.~\eqref{eq:v} and~\eqref{eq:p} in terms of their continuous Hamiltonian representations.
As opposed to the view of Ref.\cite{Toxvaerd1993} that the difference between the velocity- and position-Verlet integrators is ``only a question of notation'', we now clarify the relation between the two integrators in the following. 

i) For a system with a constant $f(x)$ independent of $x$, it is obvious that Eqs.~\eqref{eq:v} and~\eqref{eq:p} produce the same trajectories for both velocity and position. The discrete trajectories always overlap the exact solutions to the Newton's equation of motion.

ii) For the harmonic oscillator with $f(x)=-x$, Eqs.~\eqref{eq:v},~\eqref{eq:p} and~\eqref{eq:ov} can indeed produce identical positional trajectories. 
The Hamiltonians of the two integrators differ but yield the identical second-order differential equation governing the evolution of position or velocity alone. 
Once the input initial conditions are set to be: $x_0=y_0$ and $v_0 = (1-\tau^2/4) w_0$, the outputs always satisfy simple relations: $x_n = y_n$ and $v_n = (1-\tau^2/4) w_n$ for $n\geqslant 1$, as shown in Tab.~\ref{tab:1}. 
Similar relations exist whenever the velocity rather than position is focused. 
The existence of such simple relations stems from the fact that the perturbed Hamiltonians remain ``harmonic'', i.e., linear combinations of the squares of the generalized coordinate $q$ and momentum $p$.
\begin{table}[h!]\centering  \caption{The first several steps accurate up to at least six digits generated by the velocity-Verlet ($x_n$ and $v_n$) and position-Verlet ($y_n$ and $w_n$) integrators subject to $\tau = 0.2$ giving $1-\tau^2/4 = 0.99$, $x_0=y_0=0$, and $x_1=y_1=0.1$ for the harmonic and anharmonic systems.}\label{tab:1} \renewcommand{\arraystretch}{1.3}
\begin{tabular}{lcc|cccc} \hline
      & \multicolumn{2}{c}{$f(x)=-x$} & \multicolumn{4}{c}{$f(x)=-x^3$} \\
  n\quad  &  $x_n,y_n$  &  $v_n,0.99w_n$    & $x_n$        &  $v_n$      &  $y_n$         & $w_n$   \\ \hline
  0   & 0           &  0.5              & 0            &  0.5        &  0             & 0.500013 \\
  1   & 0.1         &  0.49             & 0.1          &  0.4999     &  0.1           & 0.499987 \\  
  2   & 0.196       &  0.4604           & 0.19996      &  0.499000   &  0.199930      & 0.499313 \\
  3   & 0.28416     &  0.412384         & 0.299600     &  0.495512   &  0.299481      & 0.496193 \\  \hline \end{tabular} \end{table}

iii) For a general system described by a complex force field, while the velocity-Verlet integrator still identifies with the Verlet integrator, the position-Verlet integrator does not. 
Tab.~\ref{tab:1} also numerically demonstrates that no simple relation between the position- and velocity-Verlet integrators is found for an anharmonic oscillator.

For the harmonic oscillator, the perturbed Hamiltonians corresponding to the discrete motion in Eq.~\eqref{eq:v} at the condition of $0<\tau < 2$, read\cite{Ni_Hu2024}
\begin{equation} {\cal H}_{\rm vv}(q,p,\tau|m) = \zeta_m  \left[ (1-\tau^2/4) q^2 + p^2 \right], \label{eq:h} \end{equation}
where $m$ is an arbitrary integer and the $q$- and $p$-independent coefficient $\zeta_m =2(m\pi + {\rm asin}(\tau/2))/(\tau\sqrt{4-\tau^2})$.
Given $q_0=x_0$ and $p_0 = v_0$, the discrete points, $(x_n, v_n)$ generated by Eq.~\eqref{eq:v} always lie on the continuous trajectory produced by Hamilton's canonical equations of motion:
\begin{equation} \frac{d q(t)}{dt} \equiv \dot{q} = \dfrac{\partial {\cal H}}{\partial p}; \quad \frac{d p(t)}{dt} \equiv \dot{p} = -  \dfrac{\partial {\cal H}}{\partial q}, \label{eq:ceh0} \end{equation}
that is, the solutions satisfy $q(t=n\tau) = x_n $ and $p(t=n\tau) = v_n$. 
The Hamiltonian representations are surprisingly not unique, which physically states that there are multiple continuous trajectories intersecting at all the discrete points, as shown in Fig.1 of Ref.\cite{Ni_Hu2024}.
On the other hand, the corresponding perturbed Hamiltonians for the position-Verlet integrator read\cite{Ni_Hu2024} instead
\begin{equation} {\cal H}_{\rm pv}(q,p,\tau|m) = \zeta_m  \left[ q^2 +  (1-\tau^2/4) p^2 \right]. \label{eq:hp} \end{equation}
Therefore, the continuous evolution of $(q,p)$ in the phase space follows different Hamiltonians. However, the Hamiltonians of Eqs.~\eqref{eq:h} and~\eqref{eq:hp} remain linear combinations of $q^2$ and $p^2$ with modified coefficients. 
Consequently, Eq.~\eqref{eq:ceh0}, combined with either Eq.\eqref{eq:h} or~\eqref{eq:hp}, yields the identical second-order differential equation for $q$ and $p$:
\begin{equation} \ddot{q} = - 4\zeta_m^2 (1-\tau^2/4) q;  \quad  \ddot{p} = - 4\zeta_m^2 (1-\tau^2/4) p \label{eq:cq} . \end{equation}
The continuous propagation of $q$ or $p$ alone remains the same for the two integrators. Thus, if the trajectory of the position is focused by choosing initial conditions: $q(0) = x_0 = y_0$ and $\dot{q}(0) = 2\zeta_m v_0 = 2\zeta_m (1-\tau^2/4)w_0$,  the simple relations hold at all later times:
\begin{equation} x_n = y_n = q(n\tau);\, v_n = (1-\tau^2/4)w_n =\frac{ \dot{q}(n\tau)}{2\zeta_m}, \label{eq:rx} \end{equation}
for $n=0,1,2,\cdots$. Otherwise, focusing on the trajectory of the velocity gives
\begin{equation} (1-\tau^2/4)x_n = y_n = -\frac{\dot{p}(n\tau)}{2\zeta_m};\, v_n = w_n = p(n\tau). \label{eq:rv}  \end{equation}

The validity of Eqs.~\eqref{eq:rx} and~\eqref{eq:rv} can be directly confirmed by Eqs.~\eqref{eq:v} and~\eqref{eq:p} subject to $f(x)=-x$. In fact, for the harmonic oscillator, 
Eqs.~\eqref{eq:v} and~\eqref{eq:p} with either the velocity or position eliminated, identically become the Verlet form of Eq.~\eqref{eq:ov} with $r$ interpreted correspondingly as either the position or velocity.

However, such a simple relation between the two integrators can not always be found for complex systems producing nonlinear forces.
In general, the position-Verlet integrator differs from the velocity-Verlet integrator in that it can not be reduced to the Verlet form of Eq.~\eqref{eq:ov} any more.
The evolution derived from Eq.~\eqref{eq:p} reads alternatively
\begin{multline} y_{n+2} = 2y_{n+1} - y_n + \tau^2 \\ \left[ f(y_n + \tau\, w_n/2 ) +  f(y_{n+1} + \tau\, w_{n+1}/2) \right]/2, \label{eq:p2} \end{multline}
where the last term is not equal to $f(y_{n+1})$ in general unless a linear force is considered.
Instead of evaluating the forces of Eq.~\eqref{eq:p2} at half time steps, Eqs. (4) and (5) of Ref.\cite{Toxvaerd1993} incorrectly deduced the evolution of the position from the force at the full time step. 
The difference between the two integrators for an anharmonic oscillator can be demonstrated numerically. 
Tab.~\ref{tab:1} clearly shows that, even if the initial velocities of the two integrators are chosen such that the first two positions match, difference appears soon at later times for the anharmonic oscillator giving $f(x)=-x^3$.

It should be noted that the numerical difference shown in Tab.~\ref{tab:1} is definitely not round-off errors. 
In fact, the two algorithms, when coupled to a thermostat, produce statistically different distributions (e.g.\cite{Leimkuhler_Matthews2013,Liu_Zhang2017,Liu_Shao2017}). 
Moreover, when they are used in multiple-time stepping integration\cite{Schulten1991,Tuckerman1992,Martyna_Tuckerman1996}, special stability advantages of position-Verlet over velocity-Verlet might sometimes be achieved (e.g.\cite{Batcho_Schlick2001}).

\begin{figure}[!htb]\centerline{\includegraphics[width=7.5cm]{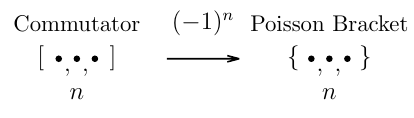} }
\caption{A schematic representation of the factor $(-1)^n$ arising from converting a $n$-th order commutator to its Poisson bracket. See the appendix for the details.}
\label{fig:c1}\end{figure}
Ideally, one might expect that the relation between any two integrators for a general system could be demonstrated through their exact Hamiltonian representations as we have done for the harmonic oscillator. 
Since the explicit solution to the discrete nonlinear dynamics is not available, we now provide a formal series expansion of the exact Hamiltonian ${\cal H}$ for general integrators of arbitrary systems. Our derivation transforms directly from a $n$-th order commutator of Liouville operators to the
corresponding Poisson bracket multiplied by $(-1)^n$ (see Fig.~\ref{fig:c1}) and thus avoids the usual complicated Lie algebra\cite{Dragt_Finn1976,Yoshida1992}.
For the anharmonic oscillator described by ${\cal H}_0 = q^4/4 + p^2/2$, the two series up to the order of $\tau^4$ reads
\begin{equation} {\cal H}_{\rm vv}(\tau,q,p) = {\cal H}_0 + \frac{\tau^2}{4}\left( q^2p^2 - \frac{q^6}{6}\right) + {\cal O}(\tau^4),  \end{equation}
for the velocity-Verlet integrator and 
\begin{equation} {\cal H}_{\rm pv}(\tau,q,p) = {\cal H}_0 + \frac{\tau^2}{4}\left( \frac{q^6}{3} - \frac{q^2p^2}{2} \right) + {\cal O}(\tau^4),   \end{equation}
for the position-Verlet integrator, respectively.
The difference between the two integrators for the nonlinear system is evident. As explained previously\cite{Ni_Hu2024}, an exact Hamiltonian representation yields the derivative of position, which is the velocity, not identical to the generalized momentum reduced by mass.
In order to reach better consistency, one might start from the original Verlet algorithm to generate positions only and then manage to obtain velocities, which differs from direct output of either velocity- or position-Verlet integrators (e.g.\cite{Gans_Shalloway2000,Jensen_Farago2013}).
On the other hand, the difference between the mass-reduced momentum and velocity arises when defining a special Hamiltonian with its kinetic energy part modified such that the propagation of position is well controlled or regulated by a bounded velocity rather than the usual derivative of $p^2/(2m)$\cite{Tuckerman_Abreu2021b}.
Recent progress along this line have demonstrated that very large outer time steps up to hundreds of femtoseconds are enabled in multiple time-stepping  molecular dynamics simulations of biological systems\cite{Tuckerman_Abreu2021b,Tuckerman_Abreu2021a}.
Therefore, recognizing the subtle difference among various integrators from the viewpoint of exact Hamiltonians  might stimulate technique advances in both single-time stepping and multiple time-stepping simulations.

In summary, we have clarified the relation between the velocity- and position-Verlet integrator numerically, analytically and formally. The present work might be useful for providing a guideline to analyze differences and similarities among many integrators in molecular dynamics simulations.

We thank the discussion with S{\o}ren Toxvaerd although we disagree with each other. We also appreciate helpful suggestions from Mark Tuckerman and stimulating discussions with Haruo Yoshida. This work was supported by NSFC (Grant No. 22273047). The authors have no conflicts to disclose.

\section*{Appendix: Jacobi identity and ${\cal H}(\tau,q,p)$ }
This appendix provides an elementary derivation of the formal series expansion of ${\cal H}(\tau,q,p)$ in the time interval $\tau$\cite{Yoshida1993},
\begin{equation} {\cal H}(\tau,q,p) = {\cal H}_0 + \sum_{n=1}^\infty \tau^n {\cal H}_n ,  \end{equation}
where ${\cal H}_0$ is the known original Hamiltonian for any given system and the first four correction terms, ${\cal H}_n$ with $n=1,2,3,4$, are to be determined explicitly and completely.
It has been known that the series is closely related to, but not exactly the same as the Baker-Campbell-Hausdorff (BCH) formula\cite{Varadarajan1984} for the product of exponential functions of non-commuting Liouville (Lie)
operators\cite{Dragt_Finn1976,Yoshida1990,Yoshida1993,Leimkuhler_Reich2005,Sanz-Serna_Calvo2018,Hairer2006}.
However, the previous derivations in several monographs\cite{Leimkuhler_Reich2005,Sanz-Serna_Calvo2018,Hairer2006} along with the used notations and definitions complicated the relation between the BCH expansion of operators and the series of ${\cal H}(\tau,q,p)$.
As emphasized by the authors, special attention must be paid to the somewhat confusing order of exponential Lie operators\cite{order_note}.
In contrast, the present transparent derivation relies on the Jacobi identity to build a simple relation between any commutator or iterated commutator of Liouville operators and its corresponding Poisson bracket, and thus leads to a straightforward expression of ${\cal H}_n$ that differs from the corresponding BCH term by a
factor of $\left(-1\right)^n$ only.

To proceed, we define the Poisson brackets of any two smooth functions of the generalized coordinates, $A(q,p)$ and $B(q,p)$, as
\begin{equation} \{ A, B \} = \frac{\partial A}{\partial q} \frac{\partial B}{\partial p} - \frac{\partial A}{\partial p} \frac{\partial B}{\partial q},  \label{eq:b} \end{equation}
and define a function-associated Liouville operator that acts on a test function $f(q,p)$ as
\begin{equation} \hat{L}_A f \equiv \hat{A} f = \{ f, A \}. \label{eq:lp} \end{equation}
According to the definition in Eq.~\eqref{eq:b}, the Liouville operator $\hat{A}$ is a first order differential operator transforming $f$ to $\{ f, A \}$.
When setting $f$ as $q$ or $p$ and regarding $A$ as a Hamiltonian, the Livouville operator $\hat{A}$ fully characterizes the evolution of the generalized coordinates in the Hamiltonian formalism
\begin{equation} \frac{dq}{dt} = \frac{\partial A}{\partial p} = \hat{A} q; \quad \frac{dp}{dt} = -\frac{\partial A}{\partial q} = \hat{A} p.\end{equation}
Hence, the exponential Livouville operator, $e^{t\hat{A}}$ formally solves the dynamics of $(q,p)$ at any time $t$.

In numerical simulations, a long time dynamics is discretized and a symplectic integrator is then used to approximate the original dynamics at a small time interval $\tau$ by means of a product of exponential Liouville operators.
For example, in a one-dimensional system described by ${\cal H}_0(q,p) = U(q) + K(p)$, typical symplectic integrators governing the discrete dynamics might take an asymmetric form
\begin{equation} e^{\tau \hat{\cal H}_0 } \simeq  e^{\tau\hat{B}} e^{\tau\hat{A}}, \label{eq:ab} \end{equation}
or a symmetric form\cite{Tuckerman1992} 
\begin{equation} e^{\tau \hat{\cal H}_0 } \simeq  e^{\tau\hat{A}/2} e^{\tau\hat{B}} e^{\tau \hat{A}/2}, \label{eq:ab2} \end{equation}
which follows from two different splits of the original Hamiltonian: ${\cal H}_0 = B + A$ or $ {\cal H}_0 = A/2 + B + A/2$.
Setting $A = U(q)$ and $B = K(p) = p^2/(2m)$, and correspondingly $\hat{A} = -\left( \partial U/\partial q\right) \partial /\partial p $ and $\hat{B} = \left( p/m\right) \partial/\partial q $ in Eqs.~\eqref{eq:ab} and~\eqref{eq:ab2} result in the symplectic Euler integrator\cite{Yoshida1993} and the well known velocity Verlet
integrator, respectively.
For the position Verlet integrator\cite{Tuckerman1992} and the adjoint Euler integrator\cite{Donnelly_Rogers2005}, $A = p^2/(2m)$ and $B = U(q) $, and thus $\hat{A} = \left( p/m\right) \partial /\partial q$ and $\hat{B} = -\left( \partial U/\partial q\right) \partial /\partial p$. 
It should be noted that different splittings or the same splitting with different choices of $\hat{A}$ and $\hat{B}$ might be converted if intermediate velocities and positions are outputted.
As pointed out by Tuckerman {\it et al.}\cite{Tuckerman1993}, the choices of $A$ and $B$ give distinct integrators but their trajectories do not diverge as a function of time. A successive application of the second-order integrator can be certainly rearranged according to 
\begin{multline} ( e^{\tau\hat{A}/2} e^{\tau\hat{B}}e^{\tau\hat{A}/2})^{n+1} = e^{\tau\hat{A}/2} e^{\tau\hat{B}} 
(e^{\tau\hat{A}} e^{\tau\hat{B}} )^n  e^{\tau \hat{A}/2} \\ = e^{\tau\hat{A}/2} ( e^{\tau\hat{B}} e^{\tau\hat{A}} )^n e^{\tau\hat{B}} e^{\tau \hat{A}/2}. \end{multline}
Therefore, trajectories produced by a second order integrator can overlap that of the first order integrator once the positions and velocities at intermediate times are specifically considered.
However, it should also be noted that, when the integrators are used for a contant-temperature simulation, the above relation might break down depending on how the thermostat is coupled. Therefore, it is crucial to distinguish among these integrators in a general case.

The $\tau$-dependent Hamiltonian ${\cal H}(\tau,q,p)$ that fully characterizes the discrete trajectory must exactly convert the product of several exponentials in Eqs.~\eqref{eq:ab} or~\eqref{eq:ab2} to the single exponential, $e^{\tau\hat{\cal H}}$. 
For the harmonic oscillator, ${\cal H}(\tau,q,p)$ can be obtained in a closed form and is found to be surprisingly non-unique for symplectic integrators even if $\tau$ is small\cite{Ni_Hu2024}. 
It is difficult to obtain any ${\cal H}(\tau,q,p)$ in an explicit form for a general nonlinear system. 
Fortunately, the ${\cal H}$-associated Liouville operator, $\hat{\cal H}$ can be easily obtained by means of the BCH expansion
\begin{equation} \hat{\cal H} = \frac{1}{\tau}\log( e^{\tau\hat{B}}e^{\tau\hat{A}} ) = \hat{\cal H}_0 + \sum_{n=1}^\infty \tau^n \hat{C}_n, \label{eq:bs} \end{equation}
and
\begin{equation} \hat{\cal H} = \frac{1}{\tau}\log( e^{\tau\hat{A}/2}e^{\tau\hat{B}}e^{\tau\hat{A}/2} ) = \hat{\cal H}_0 + \sum_{n=1}^\infty \tau^{2n} \hat{D}_{2n}, \label{eq:bs2}  \end{equation}
where $\hat{\cal H}_0 = \hat{B}+\hat{A} = \hat{A}/2 + \hat{B} + \hat{A}/2 $ by default and the coefficients $\hat{C}_n$ and $\hat{D}_n$ can all be expressed by the commutator $[\hat{A},\hat{B}]\equiv\hat{A}\hat{B} - \hat{B}\hat{A}$ and right-nested iterated commutators like
\[ [\hat{A},\hat{A},\hat{B}] = [\hat{A}, [\hat{A},\hat{B}]], [\hat{A},\hat{A},\hat{A},\hat{B}] = [\hat{A}, [\hat{A},[\hat{A},\hat{B}]]],\, \mbox{etc.} \] 
The coefficients up to the fourth order read explicitly\cite{Yoshida1990,jacobi_note}
\begin{equation}  2 \cdot \hat{C}_1 = [\hat{B},\hat{A}] , \label{eq:c1} \end{equation}
\begin{equation} 12 \cdot \hat{C}_2 = [\hat{B},\hat{B},\hat{A} ] + [\hat{A},\hat{A},\hat{B}],  \end{equation}
\begin{equation} 24 \cdot \hat{C}_3 = [\hat{B},\hat{A},\hat{A},\hat{B}] \equiv [\hat{A}, \hat{B},\hat{A},\hat{B}], \label{eq:baab} \end{equation}
\begin{equation} \begin{aligned} 720 \cdot \hat{C}_4 & = - [\hat{A},\hat{A},\hat{A},\hat{A},\hat{B}] -[\hat{B},\hat{B},\hat{B},\hat{B},\hat{A}]  \\
  & + 2 [\hat{A},\hat{B},\hat{B},\hat{B},\hat{A}] + 2 [\hat{B},\hat{A},\hat{A},\hat{A},\hat{B}] \\
  & + 6 [\hat{A},\hat{A},\hat{B},\hat{B},\hat{A}] + 6 [\hat{B},\hat{B},\hat{A},\hat{A},\hat{B}], \end{aligned} \end{equation}
\begin{equation} 24 \cdot \hat{D}_2 = 2 [\hat{B},\hat{B},\hat{A}] -[\hat{A},\hat{A},\hat{B}], \label{eq:d2} \end{equation}
and
\begin{equation} \begin{aligned} 5760 \cdot \hat{D}_4 & = 7 [\hat{A},\hat{A},\hat{A}, \hat{A}, \hat{B}] - 8 [\hat{B},\hat{B},\hat{B},\hat{B},\hat{A}] \\
 & + 16 [\hat{A},\hat{B},\hat{B},\hat{B},\hat{A}] + 16 [\hat{B},\hat{A},\hat{A},\hat{A},\hat{B}] \\
 & - 12 [\hat{A},\hat{A},\hat{B},\hat{B},\hat{A}] + 48 [\hat{B},\hat{B},\hat{A},\hat{A},\hat{B}].\label{eq:d4} \end{aligned} \end{equation}

Considering that these coefficients, $\hat{C}_n$ and $\hat{D}_{2n}$, consist of products of first order differential operators, the BCH series in the right hand side of Eqs.~\eqref{eq:bs} or~\eqref{eq:bs2} acting on a test function necessarily produce many high order differentials.
However, the sum of all second- and higher-order differentials must vanish because the Livioulle operator $\hat{\cal H}$ in the left hand side produces only the first order differentials.
In order to explore the mechanism responsible for the cancellation, we recall the Jacobi identity satisfied by the Poisson brackets of smooth functions $X(q,p)$, $Y(q,p)$ and $f(q,p)$\cite{Sanz-Serna_Calvo2018}
\begin{equation} \{ X, Y, f \} + \{Y, f, X \} + \{f, X, Y \} = 0, \label{eq:je} \end{equation}
where $\{X, Y, f\} = \{X, \{Y, f\}\}$ etc. are the right-nested iterated Poisson brackets. With the help of the $X$- and $Y$-associated Liouville operators, the first two terms of Eq.~\eqref{eq:je} can be written as
\begin{equation} \{ X, Y, f \} = -\{ \{Y, f\}, X\} = \{ \{f, Y\}, X \} = \hat{X}\hat{Y} f,   \end{equation}
and
\begin{equation} \{ Y, f, X \} = - \{ \{f, X\}, Y \} =  - \hat{Y}\hat{X} f.    \end{equation}
Therefore, the Jacobi equality essentially states that the second-order derivatives of $f$ arising from the operation of $[\hat{X},\hat{Y}]= \hat{X}\hat{Y}-\hat{Y}\hat{X}$ indeed cancel and the remaining terms with only first-order derivatives exactly identify with $- \{f, X, Y \}$, i.e.,
\begin{equation} [\hat{X},\hat{Y} ] f =  \hat{L}_{Z} f \equiv \hat{Z} f ; \quad\mbox{with}\quad Z =  - \{ X, Y \}.  \end{equation}
Similarly, typical $2$nd- and $3$rd-order commutators can be simplified explicitly according to
\begin{equation} [\hat{W}, \hat{Z} ] f = [W, X, Y] f =  \hat{S} f \end{equation}
with $S(q,p)$ being given by
\begin{equation} S = - \{ W, Z \} = \{W,\{X, Y\}\} \equiv \{W,X,Y\} ,\end{equation}
and
\begin{equation} [\hat{T}, \hat{S} ] f = [\hat{T}, \hat{W}, \hat{X}, \hat{Y} ] f   =  \hat{G} f ,\end{equation}
with $G(q,p)$ being given by
\begin{equation} G = - \{T,S\} = - \{T,W,X,Y\}, \end{equation}
respectively, for arbitrary smooth functions of $X(q,p)$, $Y(q,p)$, $W(q,p)$, and $T(q,p)$.
Clearly, as shown in Fig.~\ref{fig:c1}, any $n$th-order commutator, on account of the factor $\left(-1\right)^n$, can always be simplified exactly as its corresponding Poisson bracket-associated Liouviller operator. 
Such a simplification builds a transparent relation between $\left(-1\right)^n {\cal H}_n$ and the corresponding BCH terms of Eqs.~\eqref{eq:bs} or~\eqref{eq:bs2}. The explicit expression of ${\cal H}_n$ immediately follows Eqs.~\eqref{eq:c1} to~\eqref{eq:d4}
\begin{equation} 2 \cdot {\cal H}_1 = -\{B, A\}; \quad 24 \cdot {\cal H}_3 = -\{ B,A,A,B \}, \quad\mbox{etc.}, \end{equation} and
\begin{equation} 12 \cdot {\cal H}_2 = \{B,B,A\} - \{A,B,A\}  \quad\mbox{etc.}.  \end{equation}
for the asymmetric splitting in Eq.~\eqref{eq:ab} and 
\begin{equation} 24 \cdot {\cal H}_2 = 2\{ B, B, A \} - \{ A, A, B \} \quad\mbox{etc.}.  \end{equation}
for the symmetric splitting in Eq.~\eqref{eq:ab2}.

We have completed the derivation of the formal series expansion of ${\cal H}$ in detail.
Comparing with other derivations in the literature (e.g. Refs\cite{Leimkuhler_Reich2005,Sanz-Serna_Calvo2018,Hairer2006,order_note} and Eq.(48) to Eq.(50) of Ref.\cite{Yoshida1993}), the present derivation avoids the usual error-prone algebra, and thus allows us to easily obtain $\left(-1\right)^n {\cal H}_n$ for any large $n$ once the BCH term is given.
As a sanity check, the first few coefficients of the series expansion for the harmonic oscillator does match that from one of the explicit results ($m=0$ in Eqs.~\eqref{eq:h} and~\eqref{eq:hp}) when $\tau < 2$. 
However, the series diverges at any larger $\tau$\cite{Ni_Hu2024}. For the anharmonic oscillator described by ${\cal H}_0 = q^4/4 + p^2/2$,
\begin{multline} {\cal H}_{\rm vv}(\tau,q,p) = \frac{q^4}{4} + \dfrac{p^2}{2} + \dfrac{\tau^2}{24}\left( 6 q^2p^2- q^6 \right) + \\
               \dfrac{\tau^4}{240}(48q^4p^2-3q^8-2p^4) + {\cal O}(\tau^6) \label{eq:vv} \end{multline}
for the velocity Verlet (vv) integrator, and
\begin{multline} {\cal H}_{\rm pv}(\tau,q,p) = \dfrac{q^4}{4} + \dfrac{p^2}{2} + \dfrac{\tau^2}{24}\left( 2 q^6 -3 q^2p^2 \right)  +  \\
             \dfrac{\tau^4}{960}\left( 7p^4 - 108p^2q^4 + 48q^8 \right)+ {\cal O}(\tau^6) \label{eq:pv} \end{multline}
for the position Verlet (pv) integrator. The convergence problem of the above sequences in Eqs.~\eqref{eq:vv} to~\eqref{eq:pv} has yet to be solved.


\begin{thebibliography}{29}
\bibitem [{not()}]{note2021}%
  \BibitemOpen
  \href@noop {} {}\bibinfo {note} {Prof. S{\o}ren Toxvaerd insisted that the
  two integrators are both identical to the original verlet algorithm in recent
  private communications with us.}\BibitemShut {Stop}%
\bibitem [{jac()}]{jacobi_note}%
  \BibitemOpen
  \href@noop {} {}\bibinfo {note} {Note that Liouville operators satisfy an
  identity:
  $[\hat{A},\hat{B},\hat{C}]+[\hat{B},\hat{C},\hat{A}]+[\hat{C},\hat{A},\hat{B}]=0$,
  which, together with Eq.~\eqref{eq:baab} might be used to tranform among
  alternative BCH expansions.}\BibitemShut {Stop}%
\end{thebibliography}
\end{document}